\documentclass[a4paper]{mem}
\usepackage{natbib}
\usepackage{graphicx}

\newcommand{\be}{\begin{equation}}
\newcommand{\ee}{\end{equation}}
\newcommand{\nn}{\mbox{} \nonumber \\ \mbox{} }
\newcommand{\ba}{\begin{eqnarray}}
\newcommand{\ea}{\end{eqnarray}}

\newcommand\etal{et al.\ }
\newcommand\eg{e.g.\ }

\newcommand{\Bf}{{magnetic field\,}}
\newcommand{\Bfs}{{magnetic fields\,}}

\begin{document}
%
 \title{Polarization and  energy content of parsec  scale AGN jets}

   \author{Maxim Lyutikov \inst{1}, Vladimir Pariev \inst{2,3}, 
   Denise Gabuzda \inst{4}}

   \institute{$^1$ University of British Columbia, 
   Vancouver,  Canada  \\ 
   $^2$
University of Wisconsin Madison, Madison, USA and
 $^3$ Lebedev Physical Institute, Moscow, Russia
   \\ $^4$
     University College Cork, Cork, Ireland
             }

   \abstract{Most of  energy carried by relativistic  AGN jets
   remains undetected until hundreds of  kiloparsecs where interaction with intergalactic
   medium produces hot spots.
    The  jet's  hidden energy  is  only  partially
   dissipated at smaller  scales, from parsecs  to kiloparsecs.
   Several media may play the
   role of the ``prime mover'': ions,  pairs 
   or   large
   scale  \Bfs. 
   Analyzing VLBI  polarization structures of relativistic parsec scale
    jets we conclude that  large-scale magnetic fields can explain the
    salient polarization  properties of parsec-scale AGN jets. This  implies
    that  large-scale magnetic fields carry a non-negligible fraction
    of jet luminosity.  We also discuss the possibility that
    relativistic AGN jets may be electromagnetically (Poynting flux) dominated.
    In this case, dissipation of the toroidal magnetic field (and not fluid
    shocks) may be responsible for particle acceleration.
   \keywords{galaxies: active -- galaxies: jets -- galaxies:
               }
   }
   \authorrunning{Lyutikov}
   \titlerunning{Polarization  and  energy content of parsec  scale  AGN  jets}
   \maketitle
%

\section{Introduction}

It is generally accepted that AGN jets are produced (accelerated and 
collimated)
by electromagnetic forces originating  either near the central black hole
(\eg Blandford-Znajek mechanism, or ergosphere driving  \citep{seme04}) 
or/and
above the  accretion
disk \citep{lws87}.
At present, full 3-D general relativistic MHD simulations are beginning to probe
these mechanisms \citep{hiro04,mckin2004}. The above simulations show formation
of a  magnetically-dominated funnel, roughly in agreement with theoretical
predictions.

It is reasonable to expect that large scale fields remain in the jet as it
propagates through ISM and IGM. Such field should then show through polarization
properties of synchrotron emission. In contrast, polarization
of parsec scale jets is commonly  attributed to compression of {\it random}
\Bf at internal shocks. Motivated by this discrepancy we reconsidered 
polarization of optically thin synchrotron  emission of
parsec scale jets, trying to answer the question whether
it can be produced by large scale fields \citep{lpg04}.

\section{Polarization of  relativistic  sources}

\begin{figure*}
 \includegraphics[width=0.55 \textwidth]{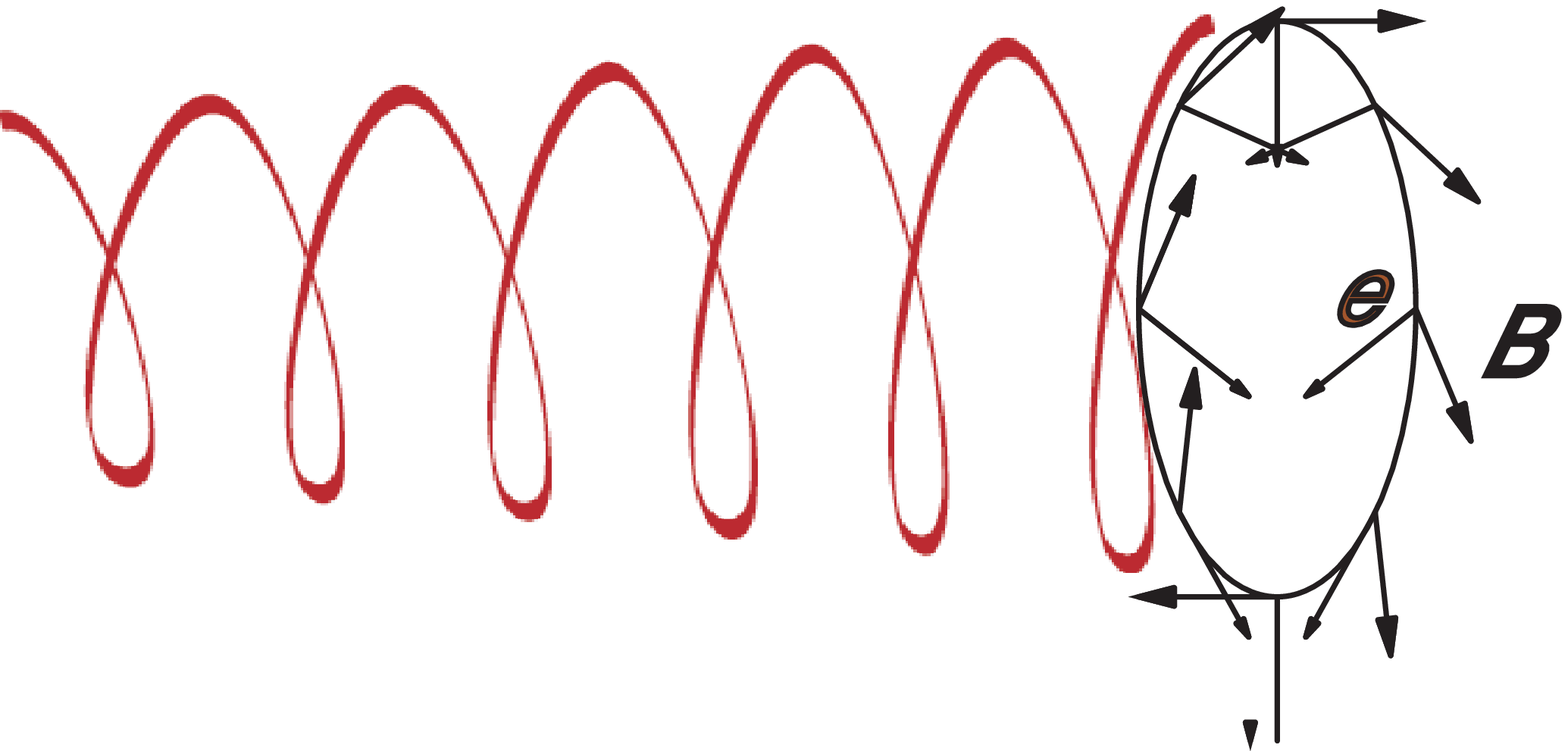}
 \includegraphics[width=0.45 \textwidth]{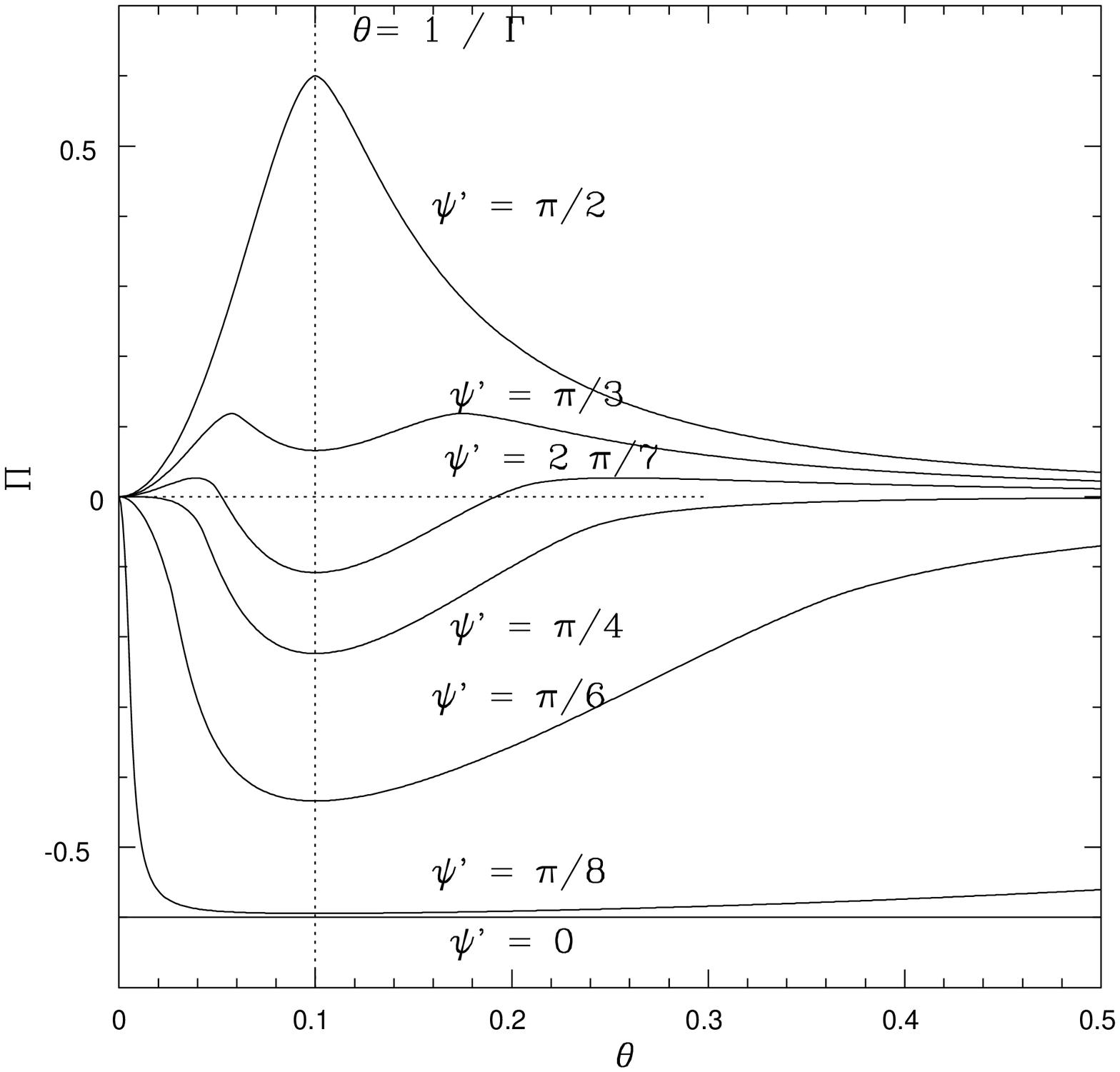}
       \caption{(a) Example of observed  electric field in the wave and
       magnetic field for  helical  field with  $\psi' = \pi/4$, $\Gamma=2$,
       $\theta_{ob} =\pi/3$.(b) Polarization fraction for unresolved cylindrical shell 
       as a function of the observer angle $\theta$ for $p=1$, $\Gamma=10$ as a function of the pitch
       angle $\psi'$. Positive values indicate polarization along the jet, negative ---
       polarization orthogonally to the jet.}
\label{AGN-Pi-bw}
   \end{figure*}

Conventionally (and erroneously for a relativistically moving plasma!), the
direction of the observed polarization  and the
associated magnetic fields are assumed to be 
 orthogonal to each other. This  is  incorrect procedure for relativistic  AGN jets since  
 relativistic boosting changes the relative strength of the magnetic
field components along and orthogonal to the line of sight, which transform
differently under the Lorentz boost, so that 
 the strength of the jet poloidal and toroidal fields measured
in the laboratory frame would be different from those measured in the jet
frame.  In addition, since the emission is boosted by the relativistic motion
of the jet material,
the polarization position angle rotates parallel to the plane
containing the velocity vector of the emitting volume ${\bf v}$ and the unit
vector in the direction to the observer ${\bf n}$, so that
{\it the observed electric field of the wave ${\hat {\bf e}}$ is not, in
general, orthogonal to the direction of unit vector
along the observed magnetic field ${\bf \hat{B}}$}
\citep{blandford79,lyu03}:
\ba &&
{ {\hat {\bf e}}} ={  {\bf n} \times {\bf q} \over
\sqrt{ q^2 - ( {\bf n} \cdot {\bf q})^2} } \mbox{,}
\nn &&
{\bf q} = {\hat{\bf B}} +
 {\bf n}  \times (  {\bf v} \times  {\hat{\bf B}})  \mbox{.}
 \label{eee}
 \ea
The angle between observed electric field (unit vector ${\hat{ \bf e}}$
and \Bf is (Fig. \ref{AGN-Pi-bw})
\be
{\hat {\bf e}} \cdot {\bf \hat{B}} =({\bf v}\times{\bf \hat{B}})\cdot({\bf n}\times{\hat {\bf e}}) \not\equiv 0
\ee
(Similar relations can be written for polarization produced at 
oblique shocks 
by compression of random \Bf.)
Thus, to reconstruct the internal structure of relativistic jets from
polarization observations, one needs to know both the polarization
{\it and } the velocity field of the jet.  Overall, plotting the direction
of the electric field  should be considered the only acceptable way
to represent polarization data for sources where relativistic motion may
be involved.

\section{Polarization from unresolved cylindrical jet with helical \Bf}

Let's the (emissivity averaged) \Bf in the jet rest frame (again properly
averaged in the case of sheared jets) be defined by toroidal field $B_\phi'$
and poloidal field $B_z'$,
so that the pitch angle $ \psi'$ {\it in the rest frame} is
$
\tan \psi'  = B_\phi'/B_z'
$. Note, that the pitch angle in the laboratory frame, 
$
\tan \psi=  \Gamma \tan \psi' $ is much larger.
One should be careful in defining what is meant by, for example, a toroidally
dominated jet. A strongly
relativistic  jet which is slightly dominated by
the  poloidal field
in its rest frame, $B_z' \geq  B_\phi'$ may be strongly toroidally dominated
in the observer frame, $ B_\phi \gg B_z$.

Qualitatively, in order for the jet polarization to be oriented along the jet axis,
the intrinsic toroidal magnetic field (in the frame of the jet) should
be of the order of or stronger than the
intrinsic poloidal field (Fig. \ref{AGN-Pi-bw}); in this case, the
highly relativistic motion of the jet implies  that, in the observer's frame,
the jet is {\it strongly } dominated by the toroidal
magnetic field $B_\phi / B_z \geq \Gamma$.

\subsection{Comparison with observations: unresolved jets}

Polarization properties expected from   jets carrying 
large scale \Bfs compare well with observations, being able to explain both the general
trends as well as possible  exceptions.
\begin{itemize}
\item  The position angle of
the electric vector of the linear polarization has a bimodal
distribution, being  oriented either parallel or perpendicular to the
jet. 
\item If an ultra-relativistic jet with $\Gamma\gg 1$
which  axis makes a small angle to the line of sight,
$\theta \sim 1/\Gamma$,
experiences a relatively small change in the direction of propagation,
velocity or pitch angle of the magnetic fields, the polarization is
likely to remain parallel or perpendicular; on the other hand, in some cases,
the degree of polarization can exhibit
large variations and the polarization position angle can experience
abrupt $90^{\circ}$ changes. This  change  is more likely to occur in jets
with flatter spectra.
\item For mildly relativistic jets, when a counter jet can be seen, the
polarization of the  counter jet is preferentially orthogonal to the axis,
unless the jet is strongly dominated by the toroidal magnetic field in
its rest frame.
\end{itemize}

\subsection{Resolved jets}

To find the total emission from a resolved jet one 
must know the distribution of
the emissivity and the internal structure of the jet, both of which are
highly uncertain functions. 
We have considered a number of force-free jet equilibria with different prescriptions for emissivities. We conclude that
\begin{itemize}
 \item In ``cylindrical shell'' type  jets (when emissivity is concentrated
 at finite radii away from the axis),  the central parts
 of the jet
 are polarized along the axis, while the outer parts are polarized
 orthogonal to it. 
 \item Contrary to the above,  a quasi-homogeneous emissivity distribution
  cannot reproduce the
  variety of position angle behavior observed, since average polarization
  remains primarily orthogonal to the jet, dominated by the central parts
  of the jet where
  the magnetic field in the co-moving frame
  is primarily poloidal (since the toroidal
  field must vanish on the axis). 
  \end{itemize}

\subsubsection{Spin of the central object}

\begin{figure*}
 \includegraphics[width=0.42 \textwidth]{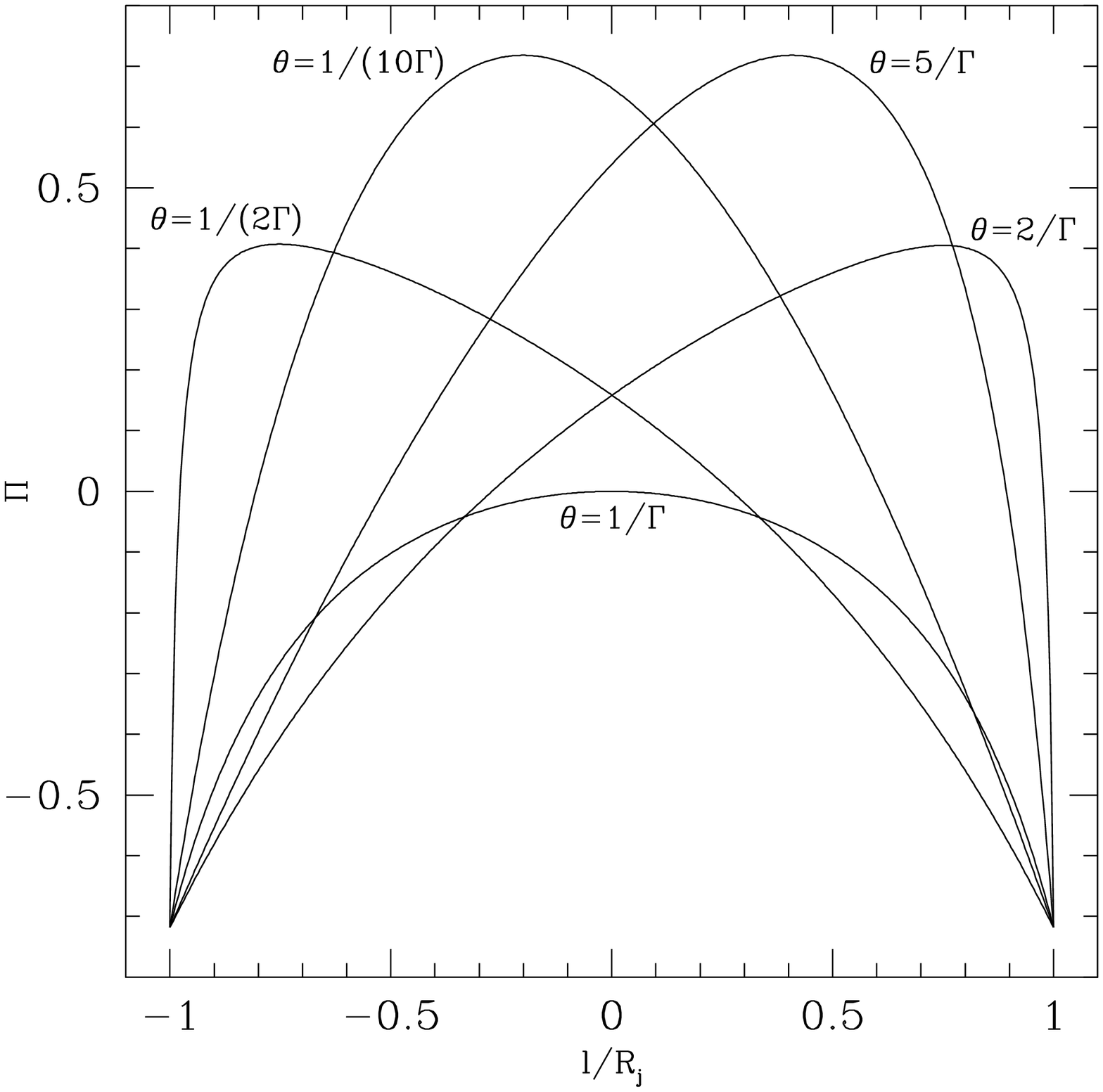}
\includegraphics[width=0.42 \textwidth]{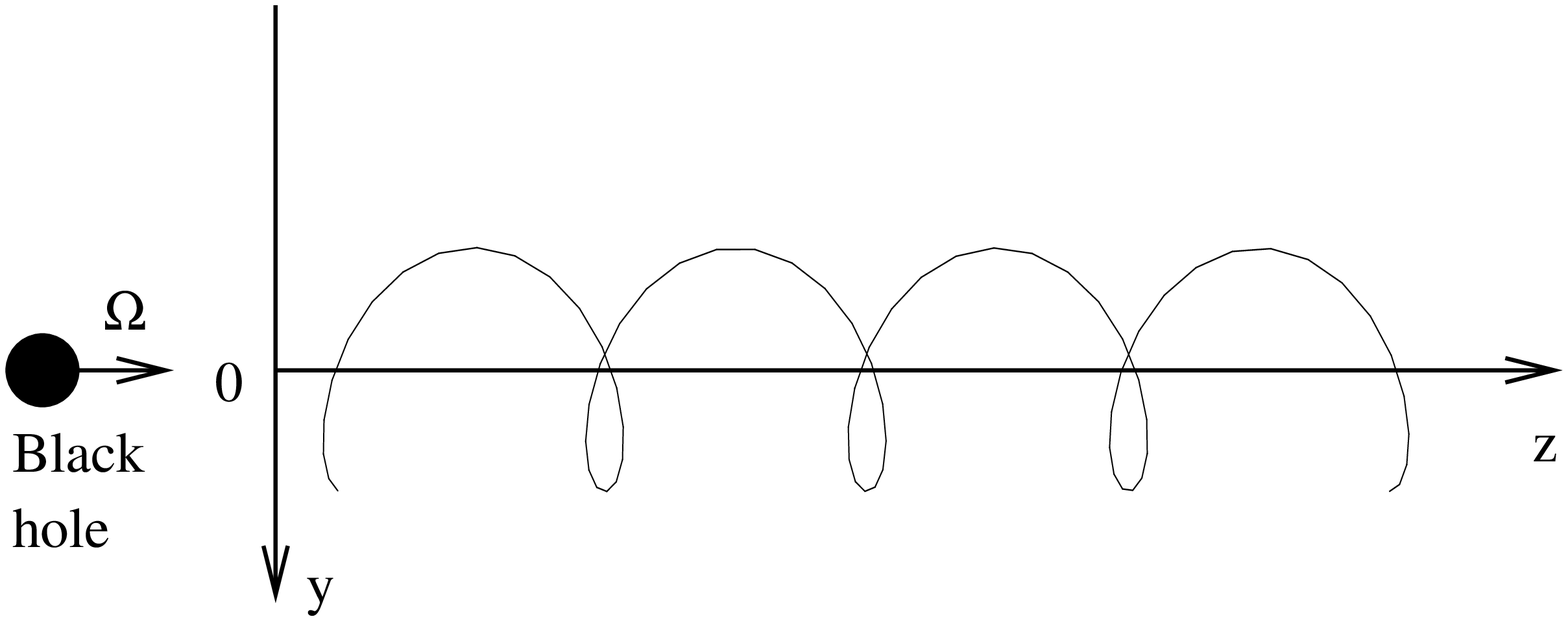}
       \caption{(a) Profiles of the polarization degree $\Pi$ for
       resolved cylindrical shells
       as a function of the distance across the jet for $\Gamma = 10$,
	$p=2.4$, $\psi' = 45^0$. (b) A ''head-on'' ($\theta \Gamma < 1$)
	view of a left-handed helical magnetic field in the reference frame co-moving
	with the jet in the positive $z$-direction.
	Polarization $\Pi$ of the lower, $y>0$, part
	is larger, i.e. it is more likely to be along the jet.
       }
\label{jetresol}
   \end{figure*}

{\it Polarization observations of resolved jets can be used to
infer the relative orientation of the  spin of the central object that
launched the jet (black hole or disk): whether it is aligned or
counter-aligned with the jet axis.} This possibility comes form the fact
that the {\it left and right-handed  helices
produce different polarization signatures, } see Figs.~\ref{jetresol}.
In order to make a distinction between the two choices, it is necessary
to determine
independently the angle at which the jet is viewed in its rest frame, which 
for 
 relativistic jets
 amounts to determining
the product $\theta \Gamma$: if $\theta \Gamma < 1$ then the jet is
viewed ``head-on,'' while the jet is viewed ``tail-on'' if
$\theta \Gamma > 1$.

\section{Conclusion}

Our   calculations indicate that
large-scale magnetic fields may be responsible for the
polarization observed in parsec-scale jets in AGNs. In order to produce the
substantial degrees of polarization that are observed in these jets, the
total energy density of the ordered component of the magnetic field must
be at least comparable to the random component.
Unless the random component is been constantly regenerated, 
large scale toroidal field will dominate over random field as the
jet expands. On the other hand, for conically expanding, constant Lorentz
factor jet the rest mass energy density falls off with radius similarly to 
toroidal \Bf energy density , $\rho c^2 , B^2 _\phi \propto r^{-2}$,
so that their ratio,
$\sigma = B^2/( 8 \pi \Gamma \rho c^2)$ remains constant.

Detectable presence of large scale \Bf
implies that $\sigma$ is not too small, $\sigma \geq 1$. 
A possibility, which remains to be proven, is that 
$\sigma \gg 1$ in the bulk of the jets, so that energy is transported along the jets
mainly in the form of the Poynting flux up to very large distances.
In this case dissipation of magnetic energy, and not fluid shocks, are responsible for particle
acceleration \citep{bla02,Par03}.

\bibliographystyle{aa}

\end{document}